\newcommand{\be}{\begin{equation}}
\newcommand{\ee}{\end{equation}}
\newcommand{\bea}{\begin{eqnarray}}
\newcommand{\eea}{\end{eqnarray}}
\newcommand{\bi}{\bigskip}

\newcommand{\noi}{\noindent}
\newcommand{\ii}{\'{\i}}

\newcommand{\Gmn}{\rm G_{\hat \mu \hat \nu}}
\newcommand{\gabmn}{\rm g^{ab}_{\hat \mu \hat \nu}}
\newcommand{\Amn}{\rm A_{\hat \mu \hat \nu}}
\documentstyle[aps,preprint]{revtex}
\begin{document}

\draft
%\twocolumn
\widetext

\title{ Is nonsymmetric gravity related to string theory?}
\author{J. A. Nieto$^a$\thanks {nie@zeus.ccu.umich.mx} and 
J. Socorro$^b$\thanks {socorro@ifug.ugto.mx}\\
$^a$Area of Superstrings, Escuela de Ciencias 
F\ii sico-Matem\'aticas \\
de la Universidad Michoacana de San Nicol\'as de Hidalgo\\
P.O. Box 749, 58000, Morelia, Michoac\'an, MEXICO\\
$^b$Instituto de F\ii sica de la Universidad de Guanajuato \\
P.O. Box E-143, 37150, Le\'on, Gto., MEXICO}

\maketitle

\begin{abstract}
In this work we raise the question whether nonsymmetric gravity and string 
theory are related. We start making the observation, that the gravitational
field $\rm g_{\mu\nu}$ and the nonsymmetric gauge field $\rm A_{\mu\nu}$
arising in the low energy limit in the string theory are exactly the
same two basic fields used in four dimensions in nonsymmetric gravity.
We argue, that this connection between nonsymmetric gravity and string theory
at the level of the gauge fields $\rm g_{\mu\nu}$ and $\rm A_{\mu\nu}$ is not,
however, reflected at the level of the corresponding associated actions. In an
effort to find a connection between such an actions we discover a new
gravitational action, which suggests an alternative version of the bosonic
string in which the target and the world-volume metrics are unified.
  
\end{abstract}

PACS numbers: {04.20.Fy, 04.50.+h, 11.25.-w}

\clearpage

\narrowtext

In this paper we make, first, a number of observations, which may suggest a 
relation between nonsymmetric gravitational theory [NGT] {\cite {11}}
and string theory (ST){\cite {22}}. Motivated by such an observations we 
compare
the action of NGT with the low energy action of the ST. We find,
that such  two actions are, in fact, unrelated. In this process we learn, that
it becomes very important to unify the  target space-time metric 
$\rm g_{\mu\nu}$ and the antisymmetric gauge field $\rm A_{\mu\nu}$ with the
world-volume metric $\rm \sqrt{-g} \, g^{ab}$ in just one object 
$\rm g^{ab}_{\mu\nu}$: the unify metric $\rm g^{ab}_{\mu\nu}$, that should 
be considered as a very general object, and that only as a particular
case $\rm g^{ab}_{\mu\nu}$ should be written as $\rm g^{ab}_{\mu\nu}= 
\sqrt{-g} \, g^{ab} \, g_{\mu\nu} + \epsilon^{ab} \, A_{\mu\nu},$  where 
$\rm \epsilon^{ab}$ is the Levi-Civita tensor in (1+1) dimensions. This
particular form of the metric $\rm g^{ab}_{\mu\nu}$ is in agreement with the
suggestion arised in the analysis of the present work.

\bi
Another interesting aspect of the present work is that the Euclidean analysis 
of the problem at hand leads us to discover a different way to understand the
complex numbers {\cite {44}} (see appendix). Moreover, in this paper, 
assuming the metric 
$\rm g^{ab}_{\mu\nu}$ as a basic fundamental object in the ST we 
conjecture an action, which presumably should be obtained in the low energy
limit of a ST based on the metric $\rm g^{ab}_{\mu\nu}$. Such an action is 
not of the nonsymmetric gravity type and 
reduces to the low energy limit to an action of the ordinary ST.

\bi
Let us start making the following observations. In ST two of the massless 
modes, besides  the dilaton, are the symmetric 
second-rank tensor $\rm g_{\mu\nu}=g_{\nu\mu},$ representing the gravitational
field potential and the antisymmetric second-rank tensor 
$\rm A_{\mu\nu}= - A_{\nu\mu}$. It turns out, that exactly the same kind of
tensor are two basic fields, used as starting point in the theory called
``nonsymmetric gravitational theory''. This observation raises the question
whether the low energy limit in ST is related to NGT. An answer to this 
question may be of physical interest, because if these two theories are not, 
in fact, related, then nonsymmetric gravity may suggest an interesting 
alternative theory to ST, which low energy behavior leads precisely to 
nonsymmetric gravity. Otherwise, it may be useful to
understand how nonsymmetric gravity arises from ordinary ST.

\bi
Another suggestion, that a relation may exist between NGT and ST comes 
from the work due to Sabbata and Gasperini {\cite {55}}, in which a 
formal coincidence is shown in the linear approximation between the Hermitian 
theory of gravity (an alternative formulation of NGT) and N=1 supergravity 
{\cite {66}}, which at the same time arises in the low energy limit of 
superstrings.

\bi
Before we proceed further let us make some general remarks about the NGT.
Its origin may be traced back to the work of Einstein and Straus {\cite {77}}.
These authors related the gauge field $\rm A_{\mu\nu}$ with the
electromagnetic field tensor. This relation, however, was not very successful.
Since the work of Einstein-Straus other alternative formulation have been 
proposed {\cite {88}}. The central idea in these alternatives is to consider 
the gauge field $\rm A_{\mu\nu}$ not as associated to the electromagnetic 
field, but rather to the gravitational field itself. Here we will consider 
the most recent version of the NGT {\cite {11}}, which presumably in the linear
approximation yields a theory free of ghost poles and tachyons and avoids the 
formation of black holes {\cite {99}} (see ref. {\cite {110}}
for some controversy about this point). So, NGT has a number of important
features, which make it a very interesting theory of gravity by
itself.

\bi
The starting point in the NGT is the descomposition of the fundamental tensor
$\rm g_{\mu\nu}$:
\be
\rm g_{\mu\nu} = g_{(\mu\nu)} + g_{[\mu\nu]},
\label {gmu}
\ee
where 
$$\rm g_{(\mu\nu)}= \frac{1}{2} \Big ( g_{\mu\nu} + g_{\nu\mu} \Big ), \,\,\,
 g_{[\mu\nu]}= \frac{1}{2} \Big ( g_{\mu\nu} - g_{\nu\mu} \Big ).$$
The action (for pure gravity) is assumed to be {\cite {11}}
\be
{\cal S}= \rm \int d^4x \, \sqrt{-g} \, g^{\mu\nu} \, R_{\mu\nu},
\label {action}
\ee
with
\be
\rm R_{\mu\nu}=  W^\beta_{\mu\nu,\beta} - \frac{1}{2} \Big (
W^\beta_{\mu\beta,\nu} + W^\beta_{\nu\beta,\mu} \Big ) 
 - W^\beta_{\alpha\nu} \, W^\alpha_{\mu\beta} + W^\beta_{\alpha\beta}
W^\alpha_{\mu\nu} ,
\label {Rmn}
\ee
where
\be
\rm W^\lambda_{\mu\nu}= \Gamma^\lambda_{\mu\nu} - \frac{2}{3} 
\delta^\lambda_\mu W_\nu.
\label {W}
\ee
In (\ref {W}) $\rm \Gamma^\lambda_{\mu\nu}$ is the connection with 
decomposition
\be
\rm \Gamma^\lambda_{\mu\nu} = \Gamma^\lambda_{(\mu\nu)} +
\Gamma^\lambda_{[\mu\nu]}.
\label {Gama}
\ee

On the other hand, the low energy limit of the ST leads to the 
action {\cite {111,112}}
\be
\rm S= \int d^{26}\, x \sqrt{-G} \Big ( {\cal R} + \frac{1}{12} 
F^{\hat \mu \hat \nu \hat \alpha}\, F_{\hat \mu \hat \nu \hat \alpha} \Big ),
\label {accion}
\ee
where $\rm G_{\hat \mu \hat \nu} = G_{\hat \nu \hat \mu} $ is the metric in
26 dimensions; ${\cal R}$ is the scalar curvature Riemann tensor defined in 
terms of
$\rm G_{\hat \mu \hat \nu}$, and
\be
\rm F_{\hat \mu \hat \nu \hat \alpha} = \partial_{[\hat \mu} \, 
A_{\hat \nu \hat \alpha]},
\label {F3}
\ee
is the completely antisymmetric field strenght defined in terms of the gauge 
field
$\rm A_{\hat \mu \hat \nu}= - A_{\hat \nu \hat \mu}$.
In (\ref {accion}) we should also consider the dilaton contribution 
$(\nabla \phi)^2$, but for the purpose of this work we droped from 
(\ref {accion}). It is important to note, that in order to derive 
(\ref {accion}) in the low energy limit of ST it is assumed, that 
$\rm G_{\hat \mu \hat \nu}$ and $\rm A_{\hat \mu \hat \nu}$ are slow varying 
background fields.

The action (\ref {accion}) may be obtained as the low energy limit 
of the quantum theory associated to the string action 
\be
\rm S= \frac{1}{2} \int d^2\, \xi \, \Big ( \sqrt{-g} \, g^{ab} \, 
G_{\hat \mu \hat \nu}
\, \partial_a \, x^{\hat \mu} \, \partial_b \, x^{\hat \nu} +
\epsilon ^{ab}\, A_{\hat \mu \hat \nu} \, \partial_a \, x^{\hat \mu}\,
\partial_b \, x^{\hat \nu} \Big ),
\label {acccion}
\ee
where $\rm g_{ab}$ is the metric on the world-surface swept out by the 
string in its evolution.

If we want to find a relation between NGT and ST we need to 
compare the two actions (\ref {action}) and (\ref {accion}). We first notice,
that in both cases the basic fields are the symmetric gauge field 
$\rm g_{(\mu\nu)}$ in the case of NGT and $\rm G_{\hat \mu \hat \nu}$ in the
case of ST, and antisymmetric gauge field $\rm g_{[\mu\nu]}$ in the
case of  NGT and $\rm A_{\hat \mu \hat \nu}$ in the case of ST.
In fact, we can make the comparison more directly, if we use the following 
definitions:
\bea
{\rm g_{(\mu\nu)}} &=&  {\rm G_{\mu\nu}} , \\ 
{\rm g_{[\mu\nu]}} &=&  {\rm A_{\mu\nu}} . 
\eea
So, at the level of the gauge fields NGT and ST differ only in the 
dimensionality of the space-time: (3+1)- dimensions in the case of NGT and
(25+1)-dimensions in the case of the bosonic ST. This observation do not
provides and essential difference since in principle one can attempt to
generalize the action (\ref {action}) to higher dimensions.  Assuming that
the basic gauge fields $\rm G_{\hat \mu \hat \nu}$ and 
$\rm A_{\hat \mu \hat \nu}$ in NGT and ST are the same, we need to concentrate
in the integrands of the two actions (\ref {action}) and (\ref {accion}).
At first sight it seems hopeless to find a connection between such 
integrands, since the integrand in (\ref {action}) looks more geometrical, than
the integrand in (\ref {accion}). However, if we success in writing 
(\ref {accion}) in a more geometrical way, we could find  real differences or
real similarities between NGT and ST. Of course, we can also use the fact, 
that
\be
\rm g_{\hat \mu \hat \nu} = G_{\hat \mu \hat \nu} + A_{\hat \mu \hat \nu},
\label {metri}
\ee
in order to develop (\ref {action}). In what follows we will follow the former
strategies.

In order to achieve our goal let us first assume, that in (\ref {metri})
$\rm G_{\hat \mu \hat \nu}$ is a real symmetric tensor and 
$\rm A_{\hat \mu \hat \nu}$ is a pure imaginary antisymmetric tensor. With
this assumption it is not difficult to show, that the metric 
$\rm g_{\hat \mu\hat \nu}$ is an Hermitian matrix, that satisfies
\be
\rm g_{\hat \mu \hat \nu}^{\dag} = g_{\hat \mu \hat \nu},
\label {hermi}
\ee
where the symbol $"\dag"$ denotes conjugate transpose.

Now, let us introduce the following definitions:
\be
\rm g_{\hat \mu \hat \nu0} \equiv G_{\hat \mu \hat \nu},
\label {def1}
\ee
and
\be
\rm g_{\hat \mu \hat \nu1} \equiv A_{\hat \mu \hat \nu}.
\label {def2}
\ee
So, with this notation the metric $\rm g_{\hat \mu \hat \nu a}$ with
a=0,1 becomes our basic object. In addition we need to introduce the metric
\be
 {\rm g_{abc}}= \left \{ \begin{array}{ll}
                 {\rm g_{ab0}}&=\frac{\rm g_{ab}}{\rm \sqrt{-g}} \\
                 {\rm g_{ab1}}&= -\epsilon_{\rm ab}.
                           \end{array}
\right.
\label {ggg}
\ee
Here $\rm g_{ab}$ is a (1+1) dimensional metric, g is the determinant of
$\rm g_{ab}$ and $\rm \epsilon_{ab}$ is a two dimensional Levi-Civita tensor
with $\epsilon_{01}=-1$.

The inverse of $\rm g_{abc}$ can be taked as
\be
 {\rm g^{abc}}= \left \{ \begin{array}{ll}
                 {\rm g^{ab0}}&= {\rm \sqrt{-g}\,  g^{ab}} \\
                 {\rm g^{ab1}}&= \epsilon^{\rm ab}.
                           \end{array}
\right.
\label {gggi}
\ee
It is interesting to observe, that the metric $\rm g_{abc}$ can be obtained by
using a vielbien field in two dimensions $\rm e_a\,^i$ and the flat metric
\be
 {\rm \eta_{ijc}}= \left \{ \begin{array}{ll}
                 {\rm \eta_{ij0}}&= {\rm  \eta_{ij}} \\
                 {\rm \eta_{ij1}}&= \epsilon_{\rm ij}.
                           \end{array}
\right.
\label {flat}
\ee
In fact, we have
\be
\rm g_{abc}= e^{-1} \, e_a\,^i \, e_b\,^j \, \eta_{ijc}
\label {gabc},
\ee
where $\rm e$ is the determinant of $\rm e_a\,^i$.

The important point here is that using (\ref {gggi}) we can define the unify
metric
\be
\rm g^{ab}_{\hat \mu \hat \nu} \equiv g^{abc} \, g_{\hat \mu \hat \nu c} .
\label {unify}
\ee
Summing over the index c in (\ref {unify}) we discover, that
\be
\rm g^{ab}_{\hat \mu \hat \nu} = \sqrt {-g} \, g^{ab} \, G_{\hat \mu \hat \nu}
 + \epsilon ^{ab} \, A_{\hat \mu \hat \nu},
\label {unify2}
\ee
where we used the definitions (\ref {def1}) and (\ref {def2}). It is not
difficult to see, that using (\ref {unify2}) the bosonic string action
(\ref {acccion}) can be written in terms of the unify metric 
$\rm g^{ab}_{\hat \mu \hat \nu},$ as
\be
\rm S= \frac{1}{2} \int d^2 \xi \, g^{ab}_{\hat \mu \hat \nu} \, 
{\partial x^{\hat \mu}
\over \partial \xi^a} \, {\partial x^{\hat \nu} \over \partial \xi^b}.
\label {ac}
\ee
This action was proposed in a previous work \cite {33}. The central idea  
was to consider the metric $\gabmn$ 
as a fundamental basic object such that, as a particular case adopt the
form (\ref {unify2}). From this point of view  either using $\beta$-function
\cite {111} or Fradkin and Tseytlin {\cite {112}} procedure one should expect 
to obtain in a low energy limit field equation derived from an action with 
the metric $\gabmn$ as a fundamental field.

What could be the form of such an action? Of course,
in the particular case, in which $\gabmn$ has the form (\ref {unify2}), 
such an 
action should reduce to the action (\ref {accion}). Just
like in ordinary ST $\rm \Gmn$ leads to the geometrical action 
$\rm \int d^{26}\, x \sqrt{-G} \, {\cal R}$, one should expect that the action,
that we are looking for has a geometrical form with $\gabmn$ as a basic 
object.  Looking things from this point of view we can first attempt to
construct the geometry behind the metric $\gabmn$ or its assocciated 
metric $\rm g_{\hat \mu \hat \nu c}= \gabmn \, g_{abc}$.

For this purpose let us define the analog of Christoffel symbols for
$\rm g_{\hat \mu \hat \nu c},$
\be
\rm \Gamma_{\hat \mu \hat \nu \hat \alpha c} = \frac{1}{2} \Big (
g_{\hat \alpha \hat \mu c, \hat \nu} +
g_{\hat \nu \hat \alpha c, \hat \mu} -
g_{\hat \nu \hat \mu c, \hat \alpha}  \Big ).
\label {simbolo}
\ee
Notice, that the position of the indices in the last term is important. It is
not difficult to see, that
\be
\rm \Gamma_{\hat \mu \hat \nu \hat \alpha} \equiv 
\Gamma_{\hat \mu \hat \nu \hat \alpha 0} = \frac{1}{2} \Big (
G_{\hat \alpha \hat \mu , \hat \nu} +
G_{\hat \alpha \hat \nu , \hat \mu} -
G_{\hat \mu \hat \nu , \hat \alpha}  \Big ),
\label {simbolo2}
\ee
is the ordinary definition of the Christoffel symbol, and that
\be
\rm F_{\hat \mu \hat \nu \hat \alpha} \equiv 
\Gamma_{\hat \mu \hat \nu \hat \alpha 1} = \frac{1}{2} \Big (
A_{\hat \alpha \hat \mu , \hat \nu} +
A_{\hat \nu \hat \alpha , \hat \mu} +
A_{\hat \mu \hat \nu , \hat \alpha}  \Big ),
\label {simbolo3}
\ee
is the completely antisymmetric field strenght considered in (\ref {accion}).
In (\ref {simbolo2}) and (\ref {simbolo3}) we used the Eqs. (\ref {def1})
and (\ref {def2}).

The curvature may be defined as
\be
\rm R_{\hat \mu \hat \nu \hat \alpha \hat \beta c} \equiv
\partial_{\hat \alpha} \, \Gamma_{\hat \nu \hat \beta \hat \mu c} -
\partial_{\hat \beta} \, \Gamma_{\hat \nu \hat \alpha \hat \mu c } +
\Gamma_{\hat \mu \hat \alpha \hat \lambda e} \, \Gamma_{\hat \nu \hat \beta
\hat \tau f} \, g^{\hat \lambda \hat \tau a} g^{efb} g_{abc} -
\Gamma_{\hat \mu \hat \beta \hat \lambda e} \, \Gamma_{\hat \nu \hat \alpha
\hat \tau f} \, g^{\hat \lambda \hat \tau a} g^{efb} g_{abc} .
 \label {curva}
\ee

General covariance implies, that the geometrical action, that we are looking 
for
should have the form
\be
{\cal S}^a = \rm \int d^{26}x\sqrt {-g} \, g^{\hat \mu \hat \alpha f} 
g^{\hat \nu \hat \beta b} \, g_{fbe}
R_{\hat \mu \hat \nu \hat \alpha \hat \beta c} \, g^{eca} ,
\label {cova}
\ee
where g is the determinant associated to $\rm g_{\hat \mu \hat \nu a}$. Of 
course, except that  ${\cal S}^a$ is a complex action (see appendix), this
action is in completly analogy to the Einstein-Hilbert action. Furthermore, we
conjecture, that it should be possible to obtain the action (\ref {cova}),
by quantization procedure  of the action (\ref {ac}) and taking the low energy limit just 
in the same way that the action (\ref {accion}) is obtained from 
(\ref {acccion}). Since the metric (\ref {unify2}) can be understood as a 
particular solution of the field equation derived from (\ref {cova}) we
should expect that (\ref {cova}) reduces to (\ref {accion}).

In fact, it is possible to show, that by choosing $\rm g_{abc}$ as 
$\rm \eta_{abc}$ and droping cubic terms of the antisymmetric gauge field 
$\Amn$ and higher in (\ref {cova}), the action ${\cal S}^0$ reduces to the 
action S, while ${\cal S}^1$ turns out to be a total derivative. The 
important point here, however, 
is that we can now compare (\ref {cova}) and (\ref {action}) since both have
a geometrical structure. We find, in fact, that both actions turn out
to be different and represent distinct theories.

It is possible, that the action (\ref {action}) may correspond to a different 
ST.  It will be curious to find the associated ST. However, one should expect
that such a new ST must be based on an action with different structure, than
the action (\ref {ac}). Anyhow, in view that NGT has unacceptable Global
Asymptotics {\cite {otro}} one should expect, that such a new ST presents
some difficulties.

\noi Note added. After the present work has been prepared for publication we
became aware, that a relation between NGT and ST has also been considered
by Moffat \cite {Mo}. It should be stressed, however, that the analysis 
developed here based in fundamental strings essentially differs, when is 
compared with the analysis of the ref. \cite {Mo} based on cosmic strings.

\appendix
\section {}
\renewcommand{\thesection}{\Alph{section}}
In \cite{44} the complex numbers were discussed from another point of view.
The central idea in such a reference is to avoid to use the imaginary
complex number i, the one, that satisfies $i^2=-1$, and to see complex
numbers as a vector $\rm A^a$ (with a=1,2) in the plane with the property
that two vectors $\rm A^a$ and $\rm B^b$ are multiplied according to the rule
\be
\rm A^a B^b  \, \omega_{abc} = C_c ,
\label {a1}
\ee
 where
\bea
\rm \omega_{ij1} = \pmatrix {1 & 0 \cr 0 & -1 }
\label {a2}
\eea
 and
\bea
\rm \omega_{ij2} = \pmatrix {0 & 1 \cr 1 & 0 } .
\label {a3}
\eea

Since the matrices (\ref {a2}) and (\ref {a3}) can be understood as reflexions
in the plane with respect to the x axis and with respect to the line 
$\rm y=x$, respectively $\rm C_1$ in (\ref {a1}) may be understood as the 
component of $\rm C_c$ obtained by making a reflexion along the x axis of the
vector $\rm B^b$ and the reflected vector been projected into the vector
$\rm A^a$. Similarly, $\rm C_2$ may be understood as the second component
of $\rm C_c$ obtained by making first one reflexion along the line $\rm y=x$ of
the vector $\rm B^b$, and then making the scalar product between the reflected
vector and $\rm A^a$. An interesting observation is that (\ref {a2}) and
(\ref {a3}) satisfy a Clifford algebra.
\begin{center}
{\LARGE Acknowledgments}
\end{center}

\noi This work was supported in part by  CONACyT Grant 4862-E9406 and by the
Coordinaci\'on de Investigaci\'on Cient\ii fica de la UMSNH.


\newpage

\begin{thebibliography}{99}
\bibitem {11} J.W. Moffat, Phys. Lett. B {\bf 355}, 447 (1995).
\bibitem {22} M. Green, J. Shwarz, and E. Witten, in {\it Supestring Theory}
	Vols 1 and 2 (Cambridge, 1987); M. Kaku, in {\it Introduction to
	Superstring} (Spring-Verlag 1988); M. Kaku, in {\it String, Conformal
	Fields, and Topology} (Spring-Verlag 1991); A.M. Polyakov, in {\it
	Gauge Fields and Strings} (Hanvood 1987).
\bibitem {44} J. A. Nieto and R. Vera, "Complex Numbers from Another Point
	of View", Esc. de Ciencias F\ii s-Mat, preprint UMSNH, (1995);
	J. A. Nieto and F. D. Mota, "Progress in Complex Numbers from
	Another Point of View", in preparation.
\bibitem {55} V, De Sabbata and M. Gasparini, {\it Hermitian Gravity with
	spin and Propagating Torsion}, in Unified Field Theories of more than
	4 Dimensions, Ed. by V. de Sabbata and E. Schmutzer (World Scientific,
	Singapore, 1983).
\bibitem {66}P. van Nieuwenhuizen, Phys. Rep. {\bf 68}, 189 (1981).
\bibitem {77} A. Einstein and E. G. Straus, Ann. Math. {\bf 47}, 731 (1946).
\bibitem {88} J. W. Moffat, \prd {\bf 19}, 3554 (1979); \prd {\bf 19}, 3562
	(1979).
\bibitem {99} J.W. Moffat, "Stellar Equilibrium and Gravitational Collapse in
	the Nonsymmetric Gravitational Theory", Preprint UTPT-95-18, 
	University of Toronto
\bibitem {110}L. M. Burko and A. Ori, \prl {\bf 75}, 2455 (1995); in "Remarks
	on the formation of black holes in nonsymmetric gravity", 
	Technion preprint TECHNION-PH-95-7.
\bibitem {111} C.G. Callan, D. Friedan, E. J. Martinec and M. J. Perry,
	Nucl. Phys. B {\bf 262}, 593 (1985).
\bibitem {112} E. S. Fradkin and A. A. Tseytlin, Phys. Lett. B {\bf 158}, 316 
	(1985).
\bibitem {33} J. A. Nieto, Il Nuov. Cim. B{\bf 110}, 225 (1995).
\bibitem {otro} T. Damour, S. Deser and J. McCarthy, Phys. Rev. D{\bf 47},
	1541 (1993).
\bibitem {Mo} J. W. Moffat, "Nonsymmetric Gravitational Theory as a String
	Theory", preprint UTPT-95-26, University of Toronto.
\end{thebibliography}
\end{document}